\documentclass{acmsiggraph}
\usepackage{amssymb}
\usepackage{graphicx}
\usepackage{multirow}
\usepackage[title]{appendix}
\usepackage[ruled,vlined]{algorithm2e}
\usepackage{array}
\usepackage{wrapfig}

\usepackage{amsmath}
\usepackage{graphicx}
\usepackage{verbatim}
\usepackage{caption}
\usepackage{epstopdf}
\usepackage[usenames,dvipsnames]{xcolor}
\usepackage{color}
\usepackage{booktabs}
\usepackage{enumitem}
\usepackage{float}

\newtheorem{theorem}{Theorem}
\newtheorem{lemma}{Lemma}

\newcommand{\norm}[1]{\left\Vert#1\right\Vert}

\newcommand{\abs}[1]{\left\vert#1\right\vert}

\newcommand{\set}[1]{\left\{#1\right\}}
\newcommand{\parr}[1]{\left (#1\right )}
\newcommand{\brac}[1]{\left [#1\right ]}
\newcommand{\tr}[1]{\mathrm{tr} #1}
\newcommand{\ip}[1]{\left \langle #1 \right \rangle }
\newcommand{\Real}{\mathbb R}

\newcommand {\1}{\mathrm{\textbf{1}}}
\newcommand{\one}{\1}

\def \S{\mathcal{S}} 
\def \T{\mathcal{T}} 

\newcommand{\fnorm}[1]{\norm{#1}_F}

\newcommand{\st}{\mathrm{s.t.}}

\newcommand{\eg}{{\it e.g.}}
\newcommand{\ie}{{\it i.e.}}

\def \etal{et al.,~}

\newcommand{\perm}{\Pi}

\renewcommand {\vec}[1]{\mathbf{#1}}

\newcommand{\RR}{\mathbb{R}}

\newcommand{\lambdaMin}{\bar \lambda_{\text{min}}}
\newcommand{\lambdaMax}{\bar \lambda_{\text{max}}}

\newcommand{\ik}{q}
\newcommand{\il}{r}
\newcommand{\ir}{s}
\newcommand{\is}{t}
\newcommand{\s}{\vec{s}}
\renewcommand{\t}{\vec{t}}
\newcommand{\SDS}{DS++}
\newcommand{\DSP}{DS+}

\title{\SDS: A Flexible, Scalable and Provably Tight Relaxation for Matching Problems}

\author{Nadav Dym\thanks{equal contribution} \and Haggai Maron\footnotemark[1] \and Yaron Lipman\\}
\affiliation{Weizmann Institute of Science}
\pdfauthor{}

\begin{document}


\maketitle

\begin{abstract}
Correspondence problems are often modelled as quadratic optimization problems over permutations. Common scalable methods for approximating solutions of these NP-hard problems are the spectral relaxation for non-convex energies and the doubly stochastic (DS) relaxation for convex energies. Lately, it has been demonstrated that semidefinite programming relaxations can have considerably improved accuracy at the price of a much higher computational cost.

We present a convex quadratic programming relaxation which is provably stronger than both DS and spectral relaxations, with the same scalability as the DS relaxation. The derivation of the relaxation also naturally suggests a projection method for achieving meaningful integer solutions which  improves upon the standard closest-permutation projection. Our method can be easily extended to optimization over doubly stochastic matrices, partial or injective matching, and problems with additional linear constraints. We employ recent advances in optimization of linear-assignment type problems to achieve an efficient algorithm for solving the convex relaxation.

We present experiments indicating that our method is  more accurate than local minimization or competing relaxations for non-convex problems. We successfully apply our algorithm to shape matching and to the problem of ordering images in a grid, obtaining results which compare favorably with state of the art methods.

We believe our results indicate that our method should be considered the method of choice for quadratic optimization over permutations.
\end{abstract}

\section{Introduction}

Matching problems, seeking some useful correspondence between two shapes or, more generally, discrete metric spaces, are central in computer graphics and vision. Matching problems are often modeled as optimization of a quadratic energy over permutations. Global optimization and approximation of such problems is known to be NP-hard \cite{QAPsurvey}.

A common strategy for dealing with the computational hardness of matching problems is replacing the original optimization problem with an easier, similar problem that can be solved globally and efficiently. Perhaps the two most common \emph{scalable} relaxations for such problems are the spectral relaxation for energies with non-negative entries \cite{leordeanu2005spectral,feng2013feature}, and the doubly stochastic (DS) relaxation for convex energies \cite{Aflalo,fiori2015spectral}. %
%
%


Our work is motivated by the recent work of \cite{Itay} who proposed  a semi-definite programming (SDP) relaxation which is provably stronger than both spectral and DS relaxations. The obtained relaxation was shown empirically to be extremely tight, achieving the global ground truth in most experiments presented. However, a major limitation was the computational cost of solving a semi-definite program with $O(n^4)$ variables. Accordingly in this paper we pursue the following question:

\textbf{Question:}
Is it possible to construct a relaxation which is stronger than  the spectral and DS relaxations, without compromising efficiency?

\begin{figure}[t]
       \includegraphics[width=1\columnwidth]{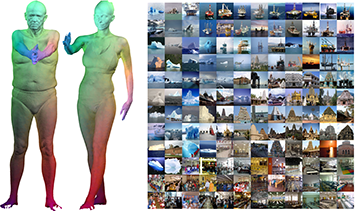}
       \caption{Our algorithm offers a flexible and scalable framework for matching metric spaces and is guaranteed to perform better than the classical spectral and doubly-stochastic relaxations. Left: non-rigid matching computed automatically between two raw scans with topological issues from the FAUST dataset \protect \cite{bogo2014faust}; Right, an automatic arrangement of natural images in a 2D grid based on deep features-based pairwise affinity. Note how similar objects are clustered together. } \label{fig:teaser}
\end{figure}

\begin{wraptable}[7]{r}{0.45\columnwidth}
        \vspace{-0.4cm}\hspace{-19pt}
        \includegraphics[width=0.5\columnwidth]{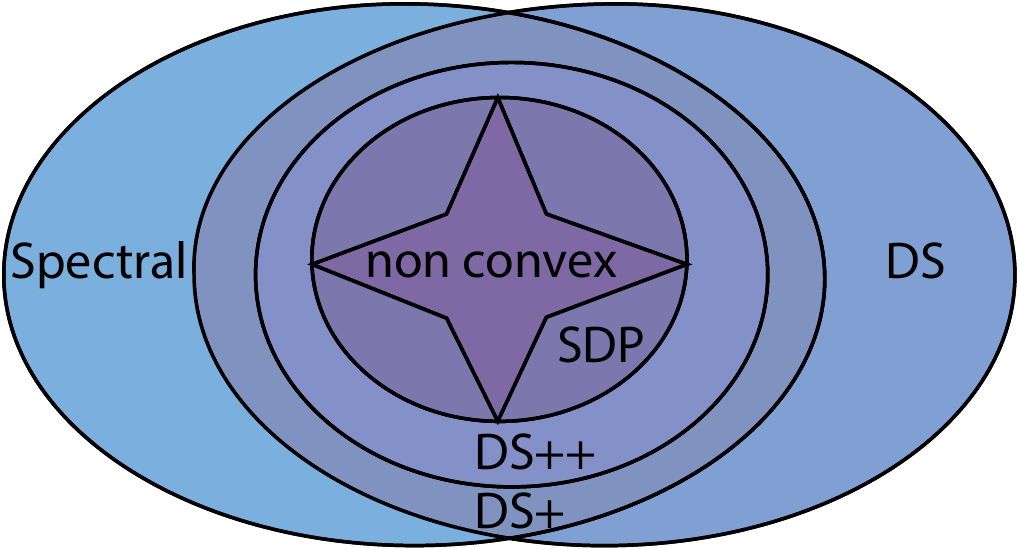}
        \vspace{-5cm}
\end{wraptable}
We give an affirmative answer to this question and show that by correctly combining the spectral and DS relaxations in the spirit of \cite{fogel2013convex} we obtain a relaxation which is provably tighter than both, and is in fact in a suitable sense exactly the intersection of both relaxations. We name this relaxation \DSP. Moreover, we observe that a refined spectral analysis leads to a significant improvement to this relaxation and a provably tighter quadratic program we name \SDS.
%
%
This relaxation enjoys the same scalability as DS and \DSP~as all three are quadratic programs with $n^2$ variables and the same number of constraints. Additional time efficiency is provided by specialized solvers for the DS relaxation such as \cite{Justin}. We note that  \SDS~is still less tight than the final expensive and accurate relaxation of \cite{Itay} yet strikes a balance between tightness and computational complexity. The hierarchy between the relaxations is illustrated in the inset and proven in section~\ref{s:proof}.

Since \SDS~is a relaxation, it is not guaranteed to output an integer solution (\ie, a permutation). To obtain a feasible permutation we propose  a homotopy-type method, in the spirit of  \cite{ogier1990neural,zaslavskiy2009path}.
This method  continuously deforms the energy functional from convex to concave,  is guaranteed to produce an integer-solution and in practice outperforms standard Euclidean projection techniques.  Essentially it provides a strategy for finding a local minima for the original non-convex problem using a good initial guess obtained from the convex relaxation.

Our algorithm is very flexible and can be applied to both convex and non-convex energies (in contrast with DS), and to energies combining quadratic and linear terms (in contrast with the spectral relaxation, which also requires energies with non-negative entries). It can also be easily modified to allow for additional linear constraints, injective and partial matching, and solving quadratic  optimization problems over the doubly stochastic matrices.
We present experiments demonstrating the effectiveness of our method in comparison to random initializations of the non-convex problem, spectral, DS, and \DSP~relaxations, as well as lifted linear-programming relaxations.

We have tested our algorithm on three applications: (i) non-rigid matching; (ii) image arrangements;  and (iii) coarse-to-fine matching. Comparison to state-of-the-art algorithms for these applications shows that our algorithm produces favorable results in comparable speed.

Our contributions in this paper are threefold: 
\begin{enumerate}
\item
We identify the optimal initial convex and concave relaxation.
\item
We show, both theoretically and experimentally that the proposed algorithm is more accurate than other popular contemporary methods. We believe that establishing a hierarchy between the various relaxation methods for quadratic matching  is crucial both for applications, and for pushing forward the algorithmic state of the art, developing stronger optimization algorithms in the future. 
\item
Lastly, we build a simple end-to-end algorithm utilizing recent advances in optimization over the doubly-stochastic matrices to provide a scalable yet accurate algorithm for quadratic matching. 
\end{enumerate}


\section{Previous work}

Many works in computer vision and graphics model correspondence problems as quadratic optimization problems over permutation matrices. In many cases these problems emerge as discretizations of isometry-invariant distances between shapes \cite{memoli2005theoretical,memoli2011gromov} . We focus here on the different methods to approximately solve these computationally hard problems.

\paragraph*{Spectral relaxation} The spectral relaxation for correspondence problems in computer vision has been introduced in \cite{leordeanu2005spectral} and has since become very popular in both computer vision and computer graphics, \eg, \cite{de2008performance,liu2009recognizing,feng2013feature,shao2013interpreting}. This method replaces the requirement for permutation matrices with a single constraint on the Frobenious norm of the matrices to obtain a maximal eigenvalue problem. It requires energies with positive entries to ensure the obtained solution is positive. This relaxation is scalable but is not a very tight approximation of the original problem. A related relaxation appears in \cite{rodola2012game}, where the variable $x$ is constrained to be non-negative with $\norm{x}_1=1$. This optimization problem is generally non-convex, but the authors suggest a method for locally minimizing this energy to obtain a sparse correspondence.

\paragraph*{DS relaxation} 
An alternative approach relaxes the set of permutations to its convex hull of doubly stochastic matrices \cite{schellewald2001evaluation}. When the quadratic objective is convex, this results in a convex optimization problem (quadratic program) which can be minimized globally, although the minimum may  differ from the global minima of the original problem. \cite{solomon2012soft} argue for the usefulness of the fuzzy maps obtained from the relaxation. For example, for symmetric shapes fuzzy maps can encode all symmetries of the shape.

  \cite{Aflalo} shows that for the convex graph matching energy the DS relaxation is equivalent to the original problem for generic asymmetric and isomorphic graphs. These results are strengthened in \cite{fiori2015spectral}. However when noise is present the relaxations of the convex graph matching energy will generally not be equivalent to the original problem \cite{journals/pami/LyzinskiFFVPS16} . Additionally, for concave energies the DS relaxation is always equivalent to the original problem \cite{gee1994polyhedral}, since minima of concave energies are obtained at extreme points. The challenge for non-convex energies is that global optimization over DS matrices is not tractable.

To achieve good initialization for local minimization of such problems, \cite{ogier1990neural,gee1994polyhedral,zaslavskiy2009path} suggest to minimize a sequence of energies $E_t$ which gradually vary from a convex energy $E_0$ to an equivalent concave energy $E_1$. In this paper we adopt this strategy to obtain an integer solution, and improve upon it by identifying the optimal convex and concave energies from within the energies $E_t$. 

The authors of \cite{fogel2013convex,FogelSiam} show that the  DS relaxation can be made more accurate by adding a concave penalty of the form $-a \norm{X}_F^2 $ to the objective. To ensure the objective remains convex they suggest to choose $a$ to be the minimial eigenvalue of the quadratic objective. We improve upon this choice by choosing $a$ to be the minimial eigenvalue \emph{over the doubly stochastic subspace}, leading to a provably tighter relaxation. The  practical advantage of our choice (\SDS) versus Fogel's choice (\DSP) is significant in terms of the relaxation accuracy as demonstrated later on.  
The observation that this choice suffices to ensure convexity has been made in the convergence proof of the softassign algorithm \cite{rangarajan1997convergence}.

\paragraph*{Optimization of DS relaxation}
Specialized methods for minimization of linear energies over DS matrices \cite{kosowsky1994invisible,cuturi2013sinkhorn,benamou2015iterative,solomon2015convolutional}  using entropic regularization and the Sinkhorn algorithm are considerably more efficient than standard linear program solvers for this class of problems. Motivated by this, \cite{rangarajan1996novel} propose an algorithm for globally minimizing quadratic energies over doubly stochastic matrices by iteratively minimizing regularized linear energies using Sinkhorn type algorithms. For the optimization in this paper we applied \cite{Justin} who offer a different algorithm for locally minimizing the Gromov-Wasserstein distance by iteratively solving regularized linear programs. The advantage of the latter algorithm over the former algorithm is its certified convergence to a critical point when applied to non-convex quadratic energies.

\paragraph*{Other convex relaxations}
Stronger relaxations than the DS relaxation can be obtained by lifting methods which add auxiliary variables representing quadratic monomials in the original variables.  This enables adding additional convex constraints on the lifted variables. A disadvantage of these methods is the large number of variables which leads to poor scalability. \cite{Itay} propose in an SDP relaxation in the spirit of \cite{zhao1998semidefinite}, which is shown to be stronger than both DS (for convex objective) and spectral relaxations, and in practice often achieves the global minimum of the original problem. However, it is only tractable for up to fifteen points. \cite{chen2015robust} use a lifted linear program relaxation in the spirit of \cite{werner2007linear,adams1994improved}. To deal with scalability issues they use Markov random field techniques \cite{kolmogorov2006convergent}  to approximate the solution of their linear programming relaxation.

\paragraph*{Quadratic assignment} Several works aim at globally solving the quadratic assignment problem using combinatorial methods such as branch and bound. According to a recent survey   \cite{QAPsurvey} these methods are not tractable for graphs with more than $30$ points. Branch and bound methods are also in need of convex relaxation to achieve lower bounds for the optimization problem. \cite{anstreicher2001new} provide a quadratic programming relaxation for the quadratic assignment problem which provably achieves better lower bounds than a competing spectral relaxation using a method which  combines spectral, linear, and DS relaxations. Improved lower bounds can be obtained using  second order cone programming \cite{xia2008second} and semi-definite programming \cite{ding2009low} in $O(n^2) $ variables. All the relaxations above use the specific structure of the quadratic assignment problem while our relaxation is applicable to general quadratic objectives which do not carry this structure and are very common in computer graphics. For example, most of the correspondence energies formulated below and considered in this paper cannot be formulated  using the quadratic assignment energy. 
\paragraph*{Other approaches for shape matching} A similar approach to the quadratic optimization approach is the functional map method (\eg,  \cite{ovsjanikov2012functional}) which solves a quadratic optimization problem over permutations and rotation matrices, typically using high-dimensional ICP provided with some reasonable initialization. Recently \cite{Maron:2016:PRV:2897824.2925913} proposed an SDP relaxation for this problem with considerably improved scalability with respect to standard SDP relaxations.

Supervised learning techniques have been successfully applied for matching specific classes of shapes in  \cite{rodola2014dense,masci2015geodesic,zuffi2015stitched,wei2016dense}. A different approach  for matching near isometric shapes is searching for a mapping in the low dimensional space of conformal maps which contains the space of isometric maps \cite{lipman2009mobius,zeng2010dense,BIM}. More information on shape matching can be found in shape matching surveys such as \cite{van2011survey}.

\section{Approach}
\paragraph*{Motivation}
 Quadratic optimization problems over the set of permutation matrices arise in many contexts. Our main motivating example is the problem of finding correspondences between two metric spaces (\eg, shapes) $(\S,d_{\S})$ and $(\T,d_{\T})$ which are related by a perfect or an approximate isometry. This problem can be modeled by uniformly sampling the spaces to obtain $\{\vec{s}_1,\ldots \vec{s}_n \} \subseteq \S $ and $\{ \vec{t}_1,\ldots,\vec{t}_n \} \subseteq \T $, and then finding the permutation $X \in \perm_n $ which minimizes an energy of the form
\begin{equation}
\label{e:matching_in_coordinates}
E(X)=\sum_{ijk\ell} W_{ijk\ell} X_{ij}X_{k\ell} + \sum_{ij} C_{ij}X_{ij}.
\end{equation}
 Here $W_{ijk\ell} $ is some penalty on deviation from isometry: If the points $\vec{s}_i,\vec{s}_k$ correspond to the points $\vec{t}_j,\vec{t}_\ell$ (resp.), then the distances between the pair on the source shape and the pair on the target shape should be similar. Therefore we  choose
\begin{equation} \label{e:W}
 W_{ijk\ell}= p ( d_{\S} (\vec{s}_i,\vec{s}_k),d_{\T} (\vec{t}_j,\vec{t}_{\ell} ))
\end{equation}
where $p(u,v)$ is some function penalizing for deviation from the set $\{(u,v) \ \vert \ u=v\} \subseteq \RR^2$. Several different choices of $p$ exist in the literature.

The linear term $C$ is sometimes used to aid the correspondence task by encouraging correspondences $\vec{s}_i \mapsto \vec{t}_j $ between points with similar isometric-invariant descriptors.
\paragraph*{Problem statement}
Our goal is to solve quadratic optimization problems over the set of permutations as formulated in \eqref{e:matching_in_coordinates}. Denoting the column stack of permutations  $X\in \Real^{n\times n}$  by the vector
$$x=[X_{11},X_{21},\ldots,X_{nn}]^T \in \RR^{n^2} $$
 leads to a more convenient phrasing of \eqref{e:matching_in_coordinates}:
\begin{subequations} \label{e:original}
        \begin{align}
      \min_X   & \quad E(X)=x^TWx+c^Tx+d \label{e:quadraticObj}\\
      \text{s.t.}  &   \quad  X \in \perm_n  \label{e:original_b}
        \end{align}
\end{subequations}
This optimization problem is non-convex for two reasons. The first is the non-convexity of $\perm_n$ (as a discrete set of matrices), and the second is that $E $ is often non-convex (if $W$ is not positive-definite).
As global minimization of  \eqref{e:original} is  NP-hard \cite{QAPsurvey} we will be satisfied with obtaining a good approximation to the global solution of \eqref{e:original} using a scalable optimization algorithm. We do this by means of a convex relaxation coupled with a suitable projection algorithm for achieving integer solutions. 
\subsection{Convex relaxation}
We formulate our convex relaxation by first considering a one-parameter family of equivalent formulations to \eqref{e:original}: observe that for any permutation matrix $X $ we have that $\fnorm{X}^2=n $. It follows that all energies of the form
\begin{equation}\label{e:Ea}
E(X,a)=E(X)-a\fnorm{X}^2+a \cdot n
\end{equation}
coincide \emph{when restricted to the set of permutations}.  Therefore, replacing the energy in \eqref{e:original} with $E(X,a)$ provides a one-parameter family of equivalent formulations. For some choices of $a$ the energy in these formulations is convex, for example, for any $a\leq \lambda_{\min}$, where $\lambda_{\min}$ is the minimal eigenvalue of $W$.

For each such equivalent formulation we consider its \emph{doubly stochastic} relaxation. That is, replacing the permutation constraint \eqref{e:original_b} with its convex-hull, the set of doubly-stochastic matrices:
\begin{subequations}\label{e:convex_relaxation}
        \begin{align}
        \min_{X}&   \quad   E(X,a)   \\
        \text{s.t.} & \quad  X\one=\one, \quad  \one^TX=\one^T \label{e:affineDS}\\
        & \quad   X \geq 0
        \end{align}
\end{subequations}
 Our goal is to pick a relaxed formulation (\ie, choose an $a$) that provides the best lower bound  to the global minimum of the original problem \eqref{e:original}. For that end we need to consider values of $a$ that make $E(X,a)$ convex and consequently turn \eqref{e:convex_relaxation} into a convex program that provide a lower bound to the global minimum of \eqref{e:original}.

Among all the convex programs described above  we would like to choose the one which provides the tightest lower bound. We will use the following simple lemma proved in Appendix \ref{app:proof}:
\begin{lemma}\label{lem:lower_bound}
For all doubly stochastic  $X$ we have $E(X,a)\leq E(X,b)$ when $a<b$, and  $E(X,a)=E(X,b)$ if and only if $X$ is a permutation.
\end{lemma}
An immediate conclusion from this lemma is that $\min_{X \in DS}E(X,a) \leq \min_{X \in DS} E(X,b) $ and so the best lower bound will be provided by the largest value of $b$ for which $E(X,b)$ is convex.

\begin{wraptable}[5]{r}{0.31\columnwidth}
        \vspace{-14 pt}\hspace*{-25 pt}
        \includegraphics[width=0.37\columnwidth]{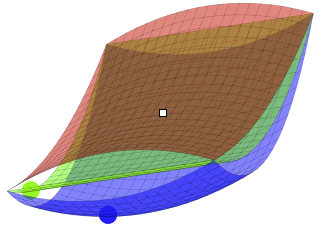}
\end{wraptable}
See for example the inset illustrating the energy graphs for different $a$ values for a toy-example: in red - the graph of the original (non-convex) energy with $a=0$; in blue the energy with $a < \lambda_{\min}$; and in green $a=\lambda_{\min}$. Note that the green graph lies above the blue graph and all graphs coincide on the corners (\ie, at the permutations). Since the higher the energy graph the better lower bound we achieve it is desirable to take the maximal $a$ that still provides a convex program in \eqref{e:convex_relaxation}. In the inset the green and blue points indicate the solution of the respective relaxed problems; in this case the green point is much closer to the sought after solution, \ie, the lower-left corner.

To summarize the above discussion: the desired $a$ is the maximal value for which $E(X,a)$ is a convex function. As noted above choosing $a=\lambda_{\min}$ in the spirit of \cite{fogel2013convex}, leads to a convex problem which we denote by \DSP. However this  is in fact not the maximal value in general. To find the maximal $a$ we can utilize the fact that $X$ is constrained to the affine space defined by the constraints \eqref{e:affineDS}: We parameterize the affine space as $x=x_0+Fz$, where $x_0$ is some permutation, $F $ is any parameterization satisfying $F^TF=I $,
and $z\in \Real^{(n-1)^2}$. Plugging this into $E(X,a)$ provides a quadratic function in $z$ of the form
$$ z^T F^T \parr{W-a I} F z + \mathrm{aff}(z)$$
where $\mathrm{aff}(z) $ is some affine function of $z$. It follows that  \eqref{e:convex_relaxation} will be convex iff  $F^T(W-aI)F$ is positive semi-definite. The largest possible $a$ fulfilling this condition is the minimal eigenvalue of $F^T W F$ which we denote by $\lambdaMin$.
Thus our convex relaxation which we name \SDS~ boils down to minimizing \eqref{e:convex_relaxation} with $a=\lambdaMin$.

\subsection{Projection}
We now describe our method for projecting the solution of our relaxation onto the set of permutations.  This method is inspired by the "convex to concave" method from \cite{ogier1990neural,gee1994polyhedral,zaslavskiy2009path}, but also improves upon these works by identifying the correct interval on which the convex to concave procedure should be applied as we now describe. 

Lemma \ref{lem:lower_bound} tells us that the global optimum of $E(X,a)$ over the doubly stochastic matrices provides an increasingly better approximation of the global optimum of the original problem \eqref{e:original} as we keep increasing $a$ even beyond the convex regime, that is $a>\lambdaMin$. In fact, it turns out that if  $a$ is chosen large enough so that $E(X,a) $ is strictly \emph{concave}, then the global optima of \eqref{e:convex_relaxation} and the global optima of the original problem over permutations are identical. This is because the (local or global) minima of strictly concave functions on a compact convex set are always obtained at the extreme points of the set. In our case, the permutations are these extreme points.

This leads to a natural approach to approximate the global optimum  of \eqref{e:original}: Solve the above convex problem with $a=\lambdaMin$ and then start increasing $a>\lambdaMin$ until an integer solution is found. We choose a finite  sequence $a_0<a_1<\ldots<a_N$, where $a_0=\lambdaMin$ and $E(X,a_N)$ is strictly concave. We begin by solving \eqref{e:convex_relaxation} with $a_0$ which is exactly the convex relaxation described above and obtain a minimizer $X_0$. We then iteratively locally minimize \eqref{e:convex_relaxation} with $a=a_i $ using as an initialization the previous solution $X_{i-1}$. The reasoning behind this strategy is that when $a_i $ and $a_{i-1}$ are close a good solution for the latter should provide a good initialization for the former, so that at the end of the process we obtain a good initial guess for the minimization of $E(X,a_N) $, which is equivalent to the original integer program. We stress that although the obtained solution may only be a local minimum, it will necessarily be a permutation.

To ensure that $E(X,a_N) $ is strictly concave we can choose any $a_N $ larger than $\lambdaMax $, which analogously to $\lambdaMin$ is defined as the largest eigenvalue of $F^T W F$. In practice we select $a_N=\lambdaMax$ which in the experiments we conducted is sufficient for obtaining integer solutions. We then took $a_i$ by uniformly sampling $[a_0,a_N] $ where unless stated otherwise we used ten samplings ($N=9$). Throughout the paper we will use the term \emph{\SDS~algorithm} to refer to our complete method (relaxation+projection) and \SDS~or \SDS~relaxation to refer only to the relaxation component. 

\begin{figure}[t]
\includegraphics[width= \columnwidth]{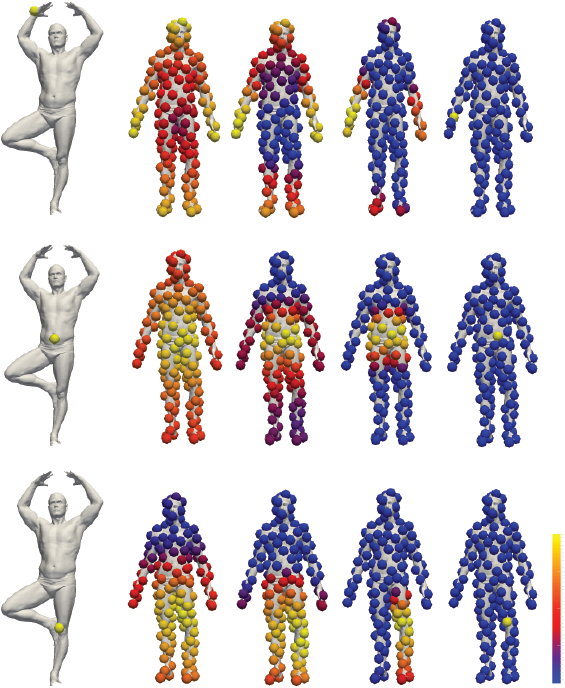}
\caption{ Visualization of the projection procedure. For each point on the source (left) a fuzzy correspondence is obtained by minimizing the convex energy (second from the left). The correspondence gradually becomes sharper as the projection procedure proceeds until the final step of minimizing a concave energy where a well defined map is obtained (right).}
\label{fig:hyst_vis}
\end{figure}

Figure~\ref{fig:hyst_vis} shows the correspondences (encoded in a specific row of $X$) obtained at different stages of the projection procedure when running our algorithm on the FAUST dataset \cite{bogo2014faust} as described below. The figure shows the correspondences obtained from optimizing $E(X,a_i)$ for $i=0,4,7,N=9$.

 Our algorithm is summarized in Algorithm~\ref{algo}:  In Section~\ref{s:implementation} we discuss efficient methods for implementing this algorithm.

\begin{algorithm}
\caption{\SDS~algorithm}
\KwIn{The energy components $W,c,d$}
\BlankLine

Compute $\lambdaMin, \lambdaMax$ of $F^TWF$\;
Choose $N+1$ uniform samples $a_0=\lambdaMin,a_1,\ldots,a_N=\lambdaMax$\;
Solve \eqref{e:convex_relaxation} with $a=a_0$ to obtain $X_0$ \;

\For{$i=1\ldots N$}
{
\BlankLine
Solve \eqref{e:convex_relaxation} with $a=a_i$ initialized from $X_{i-1} $ to obtain $X_i $ \;}
\BlankLine
\KwOut{The permutation $X_N$}
\label{algo}
\end{algorithm}

\section{Comparison with other relaxations}\label{s:proof}
The purpose of this section is to theoretically compare our relaxation with common competing relaxations. We  prove
\begin{theorem}\label{onlyTheorem}
The  \SDS~ relaxation is more accurate than \DSP, which in turn is more accurate  than the spectral and doubly stochastic relaxation.\\
The  SDP relaxation of \cite{Itay} is more accurate than all the relaxations mentioned above.
\end{theorem}

Our strategy for proving this claim is formulating all relaxations in a unified framework, using the SDP lifting technique in  \cite{Itay}, that in turn readily enables comparison of the different relaxations.

The first step in constructing SDP relaxations is transforming the original problem  \eqref{e:original} into an equivalent optimization problem in a higher dimension.The higher dimension problem is formulated over the set:
$$\perm_n^\uparrow=\set{(X,Y) \   \Big \vert \   X \in \perm_n, \quad Y=xx^T } $$
Using the identity
$$\tr{WY}=\tr{Wxx^T}=x^TWx $$
 we obtain an equivalent formulation to \eqref{e:original}:
\begin{align*}
\min_{X,Y} \quad & E(X,Y)= \tr{WY}+c^Tx+d \\
\text{s.t.}   \quad &  (X,Y) \in \perm_n^\uparrow
\end{align*}
SDP relaxations are constructed by relaxing the constraint $(X,Y)\in\perm_n^\uparrow$ using linear constraints on $X,Y$ and the semi-definite constraint $Y \succeq xx^T $.

\cite{Itay} showed that the spectral and doubly stochastic relaxations are equivalent to the following  SDP relaxations:
\begin{equation} \nonumber
\hspace{-0.1cm}
\begin{aligned}
\max & \quad E(X,Y)  \\
\mathrm{(S^\uparrow)}\quad\st        & \quad \tr\, Y =n \\
& \quad Y \succeq xx^T
\end{aligned}
\qquad \,\,\,\,\,\,\,\,
\begin{aligned}
\max & \quad  E(X,Y)\\
\mathrm{(DS^\uparrow)}\quad\st        & \quad X \in \mathrm{DS} \\
& \quad Y \succeq xx^T
\end{aligned}
\end{equation}
We note that the spectral relaxation is applicable only when $c=0$, and the DS relaxation is tractable only when the objective is convex, \ie, $W\succeq 0$. The equivalence holds under these assumptions.

Given this new formulation of spectral and DS, an immediate method for improving both relaxations is considering the \emph{Intersection-SDP}, obtained by enforcing the constraints from both $\mathrm{(DS^\uparrow)}$ and $\mathrm{(S^\uparrow)}$. The relaxation can be further improved by adding additional linear constraints on $(X,Y) $. This is the strategy followed by \cite{Itay} to achieve their final tight relaxation which is presented in Eq.~\eqref{eqn:qam_sdp} in the appendix. The main limitation of this approach is its prohibitive computational price resulting from  solving SDPs with $O(n^4)$ variables, in strong contrast to the original formulation of spectral and DS that uses only $n^2$ variables (\ie, the permutation $X$).
This naturally leads to the research question we posed in the introduction, which we can now state in more detail:

\textbf{Question:}
Is it possible to construct an SDP relaxation which is stronger than  $\mathrm{(DS^\uparrow)}$ and $\mathrm{(S^\uparrow)}$, and yet is equivalent to a tractable and scalable optimization problem with $n^2$ variables?

We answer this question affirmatively by showing that the Intersection-SDP is in fact equivalent to \DSP. Additionally \SDS~ is equivalent to a stronger SDP relaxation which includes all constraints from the \emph{Intersection-SDP}, as well as the following additional $2n^3 $ constraints: Let us write the linear equality constraints appearing in the definition of the DS matrices (\ie, \eqref{e:affineDS} ) in the form $Ax=b$. Then any $(X,Y)\in \perm_n^\uparrow$ in particular satisfies $Axx^T=bx^T$ and therefore also:
$$AY=bx^T $$
Adding these constraints to the Intersection-SDP we obtain
\begin{subequations}\label{e:ourSDP}
        \begin{align}
        \min_{X,Y}  & \quad E(X,Y) \label{e:SDPobj}\\
        \text{s.t.}    & \quad  \tr{Y}=n \label{e:SDPtrace}\\
         & \quad  X\geq 0  \label{e:SDPpositivity} \\
         & \quad Ax=b \label{e:SDPlinearEq}\\
         & \quad AY=bx^T  \label{e:SDPlinearLifted}\\
         & \quad Y \succeq xx^T \label{e:SDPSDP}
        \end{align}
\end{subequations}
Theorem~\ref{onlyTheorem} now follows from:
\begin{lemma}\label{lem:for_thm}
\begin{enumerate}
            \item The Intersection-SDP is equivalent to \DSP.
        \item The SDP relaxation in \eqref{e:ourSDP} is equivalent to  \SDS.
        \item The SDP relaxation of \cite{Itay} can be obtained by adding additional linear constraints to \eqref{e:ourSDP}.
\end{enumerate}
\end{lemma}
We prove the lemma in the appendix.

\begin{figure}
        \centering
        
        \includegraphics[width=1\columnwidth]{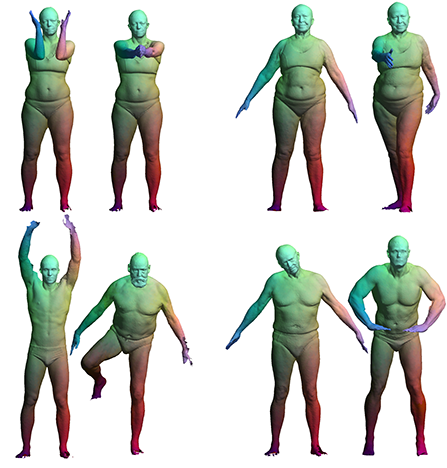}

        \caption{Typical maps obtained using our method on the FAUST dataset \protect \cite{bogo2014faust}. In each pair: left mesh is colored linearly and the computed map is used to transfer the coloring to the target, right mesh.    }
        \label{fig:FAUST_MAPS}       
\end{figure}

\begin{figure}
        \centering
        \includegraphics[width=1\columnwidth]{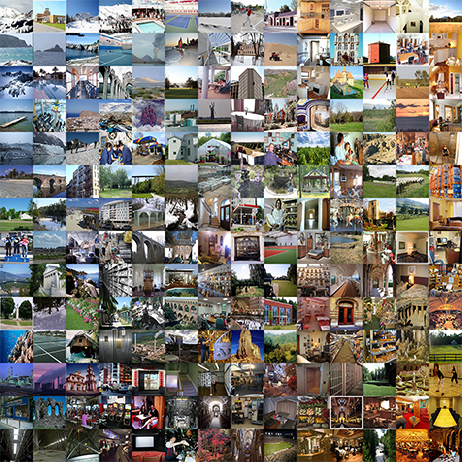}
        \label{fig:colorOrdering  }
        \caption{Image arrangement according to the mean color of images using the \SDS~algorithm. Table \protect \ref{tab:layouts} shows corresponding quantitative results. }\label{fig:colors}
\end{figure}


\section{Implementation details}\label{s:implementation}
\paragraph{Entropic regularization}
Optimization of \eqref{e:convex_relaxation} can be done using general purpose non-convex solvers such as Matlab's \emph{fmincon}, or solvers for convex and non-convex quadratic programs. We opted for the recent method of Solomon \etal \shortcite{Justin} that introduced a specialized scalable solver for local minimization of regularized quadratic functionals over the set of doubly stochastic matrices.

The algorithm of \cite{Justin}  is based on an efficient algorithm for optimizing the KL divergence
$$KL(x|y)=\ip{x,\log x}-\ip{x,\log y} $$
where $x$ is the column stack of a doubly stochastic matrix $X$ and  $y$ is some fixed positive vector. The solution for the KKT equations of this problem can be obtained analytically for $x$, up to scaling of the rows and columns, which is performed by the efficient Sinkhorn algorithm. See \cite{cuturi2013sinkhorn} for more details.

The algorithm of \cite{Justin} minimizes quadratic functionals $f(x)=x^T H x+c^T x$  (where in our case $H=W-aI$) over doubly stochastic matrices by iteratively optimizing KL-divergence problems. First the original quadratic functional is regularized by adding a barrier function $\alpha \ip{x,\log x}$ keeping the entries of $x$ away from zero to obtain a new functional
$$f_\alpha(x)=f(x)+\alpha \ip{x,\log x} $$
The parameter $\alpha$ is chosen to be some small positive number so that its effect on the functional is small.
We then define $g_\alpha(x)=\exp \parr{-\alpha^{-1}( Hx+c )}$  so that
 $$f_\alpha(x)=\alpha KL(x|g_\alpha(x))$$
 We then optimize $f_\alpha$ iteratively: In iteration $k+1$, $g_\alpha $ is held fixed at its previous value $x=x_k$, and an additional term $KL(x|x_k) $ is added penalizing large deviations of $x$ from $x_k$. More precisely, $x_{k+1}$ is defined to be the minimizer of
 $$\eta KL(x|g_\alpha(x_k))+(1-\eta)KL(x|x_k)=KL(x|g_\alpha^\eta(x_k) \odot x_k^{1-\eta})$$
where $\odot$ denotes entry-wise multiplication of vectors.
For small enough values of $\eta$,  \cite{Justin} prove that the algorithm  converges to a local minimun of $f_\alpha(x)$.

In our implementation we use $\eta=0.01$. We choose the smallest possible $\alpha$ so that all entries of the argument of the exponent in the definition of $g_\alpha$ are in $[-100,100] $. This choice is motivated by the requirement of choosing small $\alpha$ coupled with the breakdown of matlab's exponent function at around $e^{700} $. Note that this choice requires $\alpha=\alpha_k $ to update at each iteration. We find that with this choice of $\alpha$ the regularization term has little effect on the energy and we obtain final solutions which are close to being permutations. To achieve a perfect permutation we project the final solution using the $L_2$ projection. The $L_2 $ projection is computed by minimizing a linear program as described, \eg,  in \cite{zaslavskiy2009path}.

\paragraph*{Computing $\lambdaMin$ and $\lambdaMax$}
We compute $\lambdaMin$ and $\lambdaMax$ by solving two maximal magnitude eigenvalue problems: We first solve for the maximal magnitude eigenvalue of $F^T W F$. If this eigenvalue is positive then it is equal to $\lambdaMax$. We can then find $\lambdaMin$ by translating our matrix by $\lambdaMax$ to obtain a positive-definite matrix $\lambdaMax I-F^T W F$ whose maximal eigenvalue $\eta$ is related to the minimal eigenvalue of the original matrix via $\lambdaMin=\lambdaMax-\eta $.

 If the solution of the first maximal magnitude problem is negative then this eigenvalue is $\lambdaMin$, and we can use a process similar to the one described above to obtain $\lambdaMax$.

Solving maximal magnitude eigenvalue problems requires repeated multiplication of vectors $v \in \RR^{(n-1)^2} $ by the matrix $F^TWF $, where $W \in \RR^{n^2 \times n^2} $ and $F \in \RR^{n^2 \times (n-1)^2} $. If $W$ is sparse, computing $Fv$ can become a computational bottleneck. To avoid this problem, we note that $F^TWF $ has the same maximal eigenvalue as the matrix $FF^TWFF^T $ and so compute the maximal eigenvalue of the latter matrix. The advantage of this is that multiplication by the matrix $P=FF^T $ can be computed efficiently:

Since $P$ is the orthogonal projection onto $\text{Image}(F) $, we can use the identity $Pu=u-P_{\perp}u $ where $P_\perp $ is the projection onto the orthogonal complement of $\text{Image}(F) $. The orthogonal complement is of dimension $2n-1$ and therefore $ P_\perp u=F_\perp F_\perp^T u$ where $F_{\perp} \in \RR^{n^2 \times (2n-1)} $.

We solve the maximal magnitude eigenvalue problems using Matlab's function \emph{eigs}.

\section{Generalizations}\label{sec:generalizations}
\paragraph*{Injective matching}
Our method can be applied with minor changes to injective matching. The input of injective matching is  $k$ points sampled from the source shape $\S$ and $n>k $ points sampled from the target shape $\T$, and the goal is to match the $k$ points from $\S$ injectively to a subset of $\T$ of size $k$.

Matrices $X \in \RR^{k \times n}$ representing injective matching have entries in $\{0,1 \}$, and have a unique unit entry in each row, and at most one unit entry in each column. This set can be relaxed using the constraints:
\begin{subequations}
        \begin{align}
        &X \one =\one \quad , \quad \one^T X \leq \one^T\\
        &\one^TX\one=k \\
        &X \geq 0
        \end{align}
\end{subequations}

We now add a row with positive entries to the variable matrix $X$ to obtain a matrix $\bar X \in \RR^{(k+1) \times n} $ . The original matrix $X$ satisfies the injective constraints described above if $\bar X$ satisfies
\begin{align*}
& \bar X \one =(n_1-k,1,\ldots,1)^T, \quad \one^T \bar X=\one^T\\
& \bar X \geq 0
\end{align*}
These constraints are identical to the constraint defining DS, up to the value of the marginals which have no affect on our algorithm. As a result we can solve injective matching problems without any modification of our framework.

\paragraph*{Partial matching}   The input of partial matching is  $n_1,n_2$ points sampled from the source and target shape, and the goal is to match $k \leq n_1,n_2$ points from $\S$ injectively to a subset of $\T$ of size $k$. We do not pursue this problem in this paper as we did not find partial matching necessary for our applications.
However we believe our framework can be applied to such problems by adding a row and column to the matching matrix $X$.

\paragraph*{Adding linear constraints}   Modeling of different matching problems can suggest adding additional linear constraints on $X$ that can be added directly to our optimization technique. Additional linear equality constraints further decrease the dimension of the affine space $X$ is constrained to and as a result make the interval $[\lambdaMin,\lambdaMax] $ smaller, leading to more accurate optimization. We note however that incorporating linear constraints into the optimization method of \cite{Justin} is not straightforward.

\paragraph{Upsampling}  Upsampling refers to the task of interpolating correspondences between $r$ points sampled from source and target metric spaces to a match between a finer sampling of $k>>r$ source points and $n \geq k$ target points. We suggest two strategies for this problem: \emph{Limited support interpolation} and \emph{greedy interpolation}.

Limited support interpolation uses the initially matched $r$ points to rule out correspondences between the finely sampled points. The method of ruling out correspondences is discussed in the Appendix. We enforce the obtained sparsity pattern by writing $X=X_{\text{permissible}}+X_{\text{forbidden}}$, where the first matrix is zero in all forbidden entries and the second is zero in all permissible entries. We then minimize the original energy $E(X)$ only on the permissible entries, and add a quadratic penalty for the forbidden entries. That is, we minimize
$$E(X_{\text{permissible}})+\rho \norm{X_{\text{forbidden}}}_F^2 $$
choosing some large $\rho>0$. The sparsity of $X_{\text{permissible}} $ enables minimizing this energy for large $k,n$ for which minimizing the original energy is intractable.

When $k,n$ are large we use greedy interpolation. We match each source point $\s_i$ separately. We do this by optimizing over correspondences between $r+1 $ source points and $n$ target points, where the $r+1$ points are the $r$ known points and the point $s_i$. Since there are only $n-r$ such correspondences optimization can be performed globally by checking all possible correspondences.

\paragraph*{Optimization over doubly stochastic matrices} Our main focus was on optimization problems over permutations. However in certain cases the requested output from the optimization algorithm may be a doubly stochastic matrix and not a permutation. When the energy $E$ is non convex this still remains a non-convex problem. For such optimization problems our method can be applied by taking samples $a_i$ from the interval $[\lambdaMin,0]$, since minimization of \eqref{e:convex_relaxation} with $a=0$ is the problem to be solved while minimization of \eqref{e:convex_relaxation} with $a=\lambdaMax$ forces a permutation solution.

\begin{figure}[t!]
        \includegraphics[width=\columnwidth]{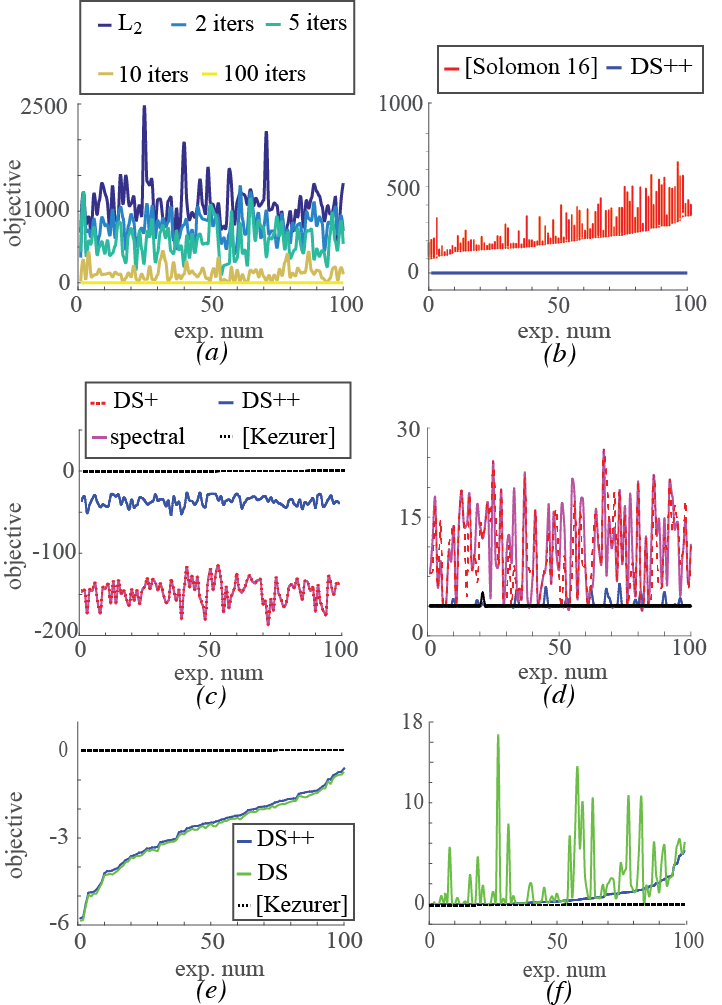}
        \caption{Evaluation of our algorithm. (a) compares the $L_2$ projection with our projection. Even with only two iterations our projection improves upon the $L_2$ projection. Additional iterations yield better accuracy at the price of time complexity. (b)  compares minimization of the Gromov-Wasserstein distance with our algorithm and  \protect\cite{Justin} with $1000$ random initializations. In all cases we attain a lower objective value. The second row compares lower bounds (c) and upper bounds (d) obtained by the \SDS~algorithm, \DSP, spectral,  and \protect\cite{Itay}. As predicted by Theorem~\ref{onlyTheorem} our lower bounds and upper bounds are outperformed by \protect\cite{Itay} who are able to attain the ground truth in these cases, but improve upon those of the remaining methods. The third row compares lower bounds (e) and upper bounds (f) obtained by the \SDS~algorithm, DS and \protect\cite{Itay} for the convex graph matching functional. The lower bound of the \SDS~algorithm modestly improves DS's, while  the upper bounds substantially improves the upper bounds of  DS's $L_2$ projection. }
        \label{fig:evaluation}
        \vspace{-0.5cm}
\end{figure}
\begin{figure}
        \includegraphics[width=\columnwidth]{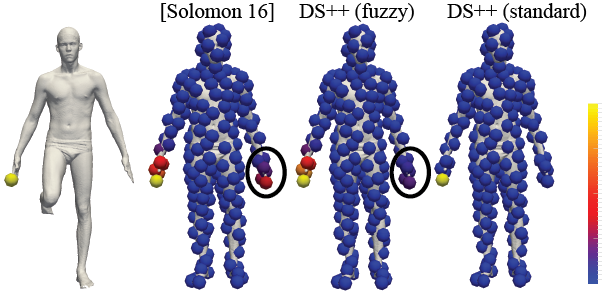}
        \caption{Optimization over fuzzy maps using \protect \cite{Justin} and the \SDS~algorithm as described in Section \protect \ref{sec:generalizations}.  The best fuzzy map obtained by \protect \cite{Justin} with $1000$ random initializations is less accurate than our fuzzy map (middle), as our map gives lower probability to mapping the right hand of the source to the left hand of the target. See also Figure \ref{fig:evaluation} (b). The rightmost image shows the sharp map obtained by the standard \SDS~algorithm.}
        \label{fig:justin_qualitative}
\end{figure}
\section{Evaluation}
In this section we evaluate our algorithm and compare its performance with relevant state of the art algorithms. We ran all experiments on the 100 mesh pairs from the FAUST dataset \cite{bogo2014faust} which were used in the evaluation protocol of \cite{chen2015robust}.

\paragraph*{Comparison with \cite{Justin}}
In figure~\ref{fig:evaluation}(b) we compare our method for minimizing non-convex functionals with the local minimization algorithm of \cite{Justin}. Since this method is aimed at solving non-convex functionals over doubly-stochastic matrices, we run our algorithm using samples in $[\lambdaMin,0] $ as explained in Section~\ref{sec:generalizations}. We  sample $200$ points from each mesh using farthest point sampling \cite{eldar1997farthest}, and  optimize the Gromov-Wasserstein (GW) functional advocated in \cite{Justin}, which amounts to choosing $p$ from \eqref{e:W} to be $p(u,v)=(u-v)^2$.
As local minimization depends on initialization we locally minimize $1000$ times per mesh pair, using $1000$ different random initializations. The initializations are obtained by randomly generating a positive matrix in $\RR^{200 \times 200}$ with uniform distribution, and projecting the result onto the doubly stochastic matrices using the Sinkhorn algorithm. As can be seen in the figure our algorithm, using only ten iterations, was more accurate than all the local minima found using random initializations. As a baseline for comparison we note that the difference in energy between randomly drawn permutations and our solution was around $5000$, while the difference in energy shown in the graph is around $500$. 
Figure~\ref{fig:justin_qualitative} visualizes the advantages of the fuzzy maps obtained by our algorithm in this experiment  over the best of the $1000$ random maps generated by \cite{Justin}.

\paragraph*{Projection evaluation} In figure~\ref{fig:evaluation}(a) we examine how the result obtained from our projection method is influenced by the number of points $N$ sampled from $[\lambdaMin,\lambdaMax]$. We compared the behavior of our relaxation with several different choices of $N$ as well as with the standard $L_2 $ projection onto the set of permutations. As expected, our projection is always better that the $L_2$ projection, and the projection improves as the number of samples is increased.

\paragraph*{Comparison with other relaxations}
We compare our method with other relaxation based techniques. In figure~\ref{fig:evaluation} (c)-(d) we compare our relaxation with the spectral relaxation, the \DSP~ relaxation, and the SDP relaxation of \cite{Itay}. In this experiment  the energy we use is non-convex so DS is not applicable.

 We sampled $10$ points
from both meshes, and minimized the (non-convex)  functional selected by \cite{Itay}, which amounts to choosing $p$ from \eqref{e:W} to be
$$p(u,v)=-\exp\left( \frac{-(u-v)^2}{\sigma^2} \right) $$
We choose the parameter $\sigma=0.2$. For the minimization we  used all four relaxations, obtaining a lower bound for the optimal value, Figure \ref{fig:evaluation} (c). We then projected the solutions obtained onto the set of permutations, thus obtaining an upper bound, Figure \ref{fig:evaluation} (d). For methods other than the \SDS~algorithm we used the  $L_2$ projection. In all experiments the upper and lower bounds provided by the SDP relaxation of \cite{Itay} were identical, thus proving that the SDP relaxation found the globally optimal solution.  Additionally, in all experiments the upper bound and lower bound provided by our relaxation were superior to those provided by the spectral method, and  our projection attained the global minimum in approximately $80\%$ of the experiments in contrast to $11\%$  obtained by the $L_2$ projection of the spectral method. The differences between the spectral relaxation and the stronger \DSP~relaxation were found to be negligible.

 In figure~\ref{fig:evaluation}(e)-(f) we perform the same experiment, but now we minimize the convex graph matching functional  $E(X)=\fnorm{AX-XB}^2 $ from \cite{Aflalo} for which the classical DS relaxation is applicable. Here again the ground truth is achieved by the SDP relaxation. Our relaxation can be seen to modestly improve the lower bound obtained by the classical DS relaxation, while our projection method substantially improves upon the standard projection.

\section{Applications}\label{sec:apps}
We have tested our method for three applications: non-rigid shape matching, image arrangement, and coarse-to-fine matching.

\paragraph*{Non-rigid matching}
\begin{figure}[t]
        \centering
        
        \includegraphics[width=1\columnwidth]{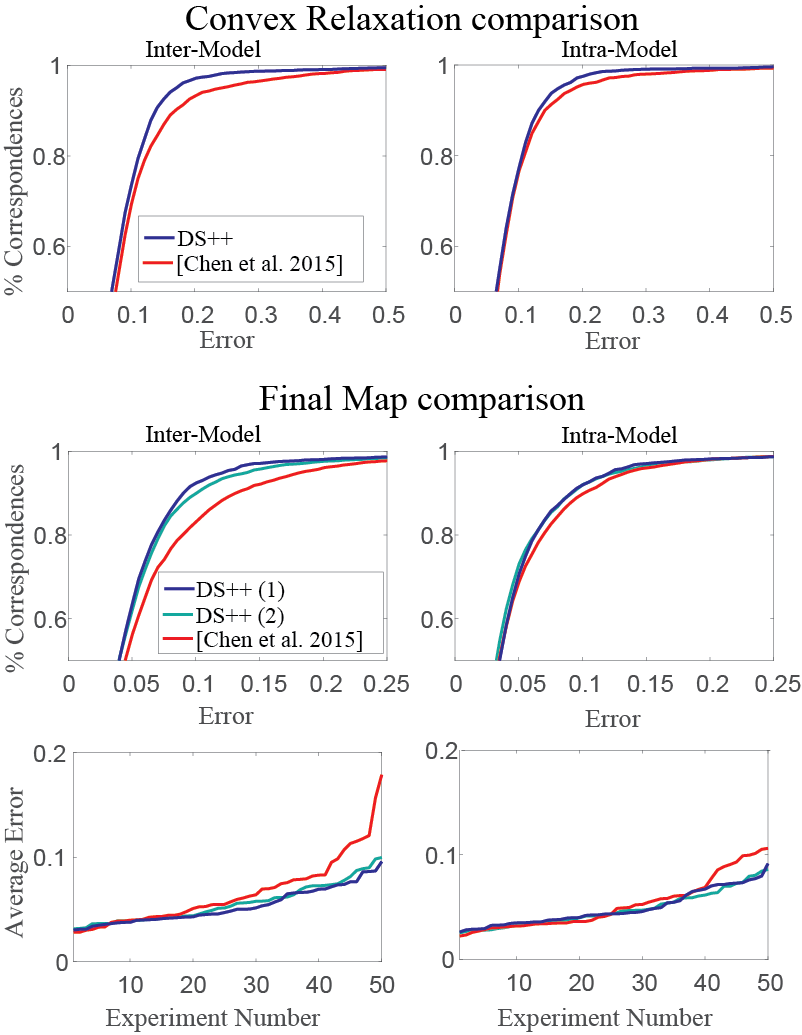}
        
        \caption{ Non-rigid matching. Cumulative and average errors achieved on the FAUST dataset \protect\cite{bogo2014faust} by the \SDS~algorithm compared to  \protect\cite{chen2015robust}. Top row compares only the convex relaxation part of both methods; bottom two rows compare final maps after upsampling. \SDS (1) uses our upsampling method and \SDS (2) uses the upsampling method of \protect \cite{chen2015robust}.}
        \label{fig:FAUST_GRAPH}
        
\end{figure}

We evaluated the performance of our algorithm for non-rigid matching on the FAUST dataset  \cite{bogo2014faust}. We compared to \cite{chen2015robust} which demonstrated superb state of the art results on this dataset (for non learning-based methods). For a fair comparison we used an identical pipeline to \cite{chen2015robust}, including their isometric energy modeling and extrinsic regularization term. We first use the \SDS~algorithm to match $n=160, k=150$ points, then upsampled to $n=450, k=420$ using limited support interpolation and to $n=5000, k=1000$ using greedy interpolation, as described in Section \ref{sec:generalizations}; the final point resolution is as in \cite{chen2015robust}.

Figure~\ref{fig:FAUST_GRAPH} depicts the results of the  \SDS~algorithm and \cite{chen2015robust}. As can be read from the graphs, our algorithm compares favorably in both the inter and intra class matching scenarios in terms of cumulative error distribution and average error. These results are consistent for both the convex relaxation part (top row) and the upsampled final map (middle row); The graphs show our results both with our upsampling as described above (denoted by \SDS (1)) and the results of combining our relaxation with the upsampling of \cite{chen2015robust} (\SDS (2)). We find \SDS(1) to be better on the inter class, and \SDS(2) is marginally better on the intra class.
The error is calculated on a set of 52 ground truth points in each mesh as in \cite{Maron:2016:PRV:2897824.2925913}. Figure \ref{fig:teaser} (left), and \ref{fig:FAUST_MAPS} show typical examples of maps computed using the \SDS~algorithm in this experiment.

\paragraph*{Image arrangement}

\begin{table*}
	\centering
	\begin{tabular}{lcccccccc}
		dataset & \multicolumn{1}{l}{ feature} & \multicolumn{1}{l}{improvement} & \multicolumn{1}{l}{rand  average} & \multicolumn{1}{l}{Fried mean} & \multicolumn{1}{l}{our mean} & \multicolumn{1}{l}{functional} & \multicolumn{1}{l}{swaps?} & \multicolumn{1}{l}{grid size} \\
		\midrule
		\midrule
		\textit{Random colors} & color & \textbf{28.33\%} & 0.478 & 0.259 & \textbf{0.198} & Fried  & no    & 12 \\
		\textit{Random colors} & color & \textbf{8.86\%} & 0.478 & 0.219 & \textbf{0.197} & Fried  & yes   & 12 \\
		\textit{Random colors} & color & \textbf{3.46\%} & 0.478 & 0.219 & \textbf{0.211} & GW    & yes   & 12 \\
		\hline             \textit{SUN dataset } & color & \textbf{2.05\%} & 0.581 & 0.244 & \textbf{0.237} & Fried  & no    & 10 \\
		\textit{SUN dataset} & color & \textbf{0.57\%} & 0.581 & 0.225 & \textbf{0.223} & Fried  & yes   & 10 \\
		\hline               \textit{SUN dataset} & deep feature object & \textbf{55.97\%} & 0.433 & 0.345 & \textbf{0.295} & Fried  & no    & 14 \\
		\textit{SUN dataset} & deep feature object & \textbf{6.31\%} & 0.433 & 0.300 & \textbf{0.292} & Fried  & yes   & 14 \\
		\hline
		\textit{LFW} & deep feature face & \textbf{50.70\%} & 0.422 & 0.355 & \textbf{0.320} & Fried  & no    & 14 \\
		\textit{LFW} & deep feature face & \textbf{2.81\%} & 0.422 & 0.321 & \textbf{0.318} & Fried  & yes   & 14 \\
		\hline
		\textit{Illumination} & Raw $L_2$ distance & \textbf{59.08\%} & 0.509 & 0.320 & \textbf{0.208} & Fried  & no    & 10 \\
		\textit{Illumination} & Raw $L_2$ distance & \textbf{9.94\%} & 0.509 & 0.232 & \textbf{0.204} & Fried  & yes   & 10 \\
		\textit{Illumination} & Raw $L_2$ distance & \textbf{13.70\%} & 0.527 & 0.273 & \textbf{0.238} & Fried  & yes   & 10 \\
		\textit{Illumination} & Raw $L_2$ distance & \textbf{10.65\%} & 0.518 & 0.259 & \textbf{0.231} & Fried  & yes   & 10 \\
		\bottomrule
	\end{tabular}%
	\caption{Image arrangement comparison. We compare \SDS\ to \protect \cite{fried2015isomatch} in arranging different sets of images in a grid with different affinity measures between images; see text for more details. }
	\label{tab:layouts}%
\end{table*}%

The task of arranging image collections in a grid has received increasing attention in recent years \cite{quadrianto2009kernelized,strong2014self,fried2015isomatch,carrizosa2016visualizing}. Image arrangement is an instance of metric matching: the first metric space is the collection of images and a dissimilarity measure defined between pairs of images; and the second, target metric space is a 2D grid (generally, a graph) with its natural Euclidean metric.

\cite{fried2015isomatch} suggested an energy functional for generating image arrangements, which are represented by a permutation matrix $X$. Their choice of energy functional was supported by a user study. This energy functional is:
\begin{equation}\label{e:fried}
E(X)=\min_{c>0} \sum_{ijkl} \abs{c\cdot d_{ik}-d'_{jl}}X_{ij}X_{kl}
\end{equation}
where $d,d'$ are the distance measures between images and grid points respectively, and  $c$ is the unknown scale factor between the two metric spaces.
\cite{fried2015isomatch} suggested a two step algorithm to approximate the minimizer of the above energy over the set of permutations: The first step is a dimensionality reduction,  and the second is  linear assignment to a grid according to Euclidean distances. Fried \etal demonstrated significant quantitative improvement over previous state of the art methods.

We perform image arrangement by using an alternative method for optimizing the energy  \eqref{e:fried}. We fix $c$ so that the mean of $d$ and $d'$ are the same,  which leads to a quadratic matching energy which we  optimize over  permutations using the \SDS~algorithm.


Table~\ref{tab:layouts} summarizes quantitative comparison of the \SDS~algorithm and \cite{fried2015isomatch} on a collection of different image sets and dissimilarity measures. Each row shows the mean energies over 100 experiments  of Fried \etal \SDS, and random assignments which provide a baseline for comparison; in each experiment we randomized a subset of  images from the relevant set of images and generated image arrangements using the two methods.  \cite{fried2015isomatch} suggested an optional post processing  step in which random swaps are generated and applied in case they reduce the energy; our experiment measures the performance of both algorithms with and without random swaps. The first set of experiments tries to arrange random colors in a grid. The second set of experiments uses the mean image color for images form the SUN database \cite{xiao2010sun}. The third set uses the last layer of a deep neural network
trained for object recognition \cite{chatfield2014return} as image features, again for images in the SUN dataset. The fourth set of experiments organizes randomly sampled images from the Labeled Faces in the Wild (LFW) dataset \cite{huang2007labeled} according to similar deep features taken from the net trained by
\cite{parkhi2015deep}. For the last experiment, we rendered three 3D models from the SHREC07 dataset \cite{Shrec} from various illumination directions and ordered them according to the raw $L_2$ distance between pairs of images.
\begin{wrapfigure}[15]{r}{0.5\columnwidth}
	\includegraphics[width=0.5\columnwidth]{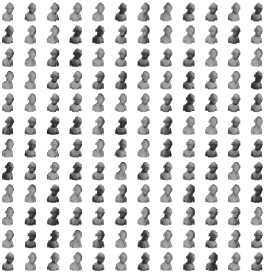}
	\vspace{-0.5cm}
	\caption{Random lighting.}
	\label{fig:randomLighting}
\end{wrapfigure}

\begin{figure*}
	\centering
	
	\includegraphics[width=1.99\columnwidth]{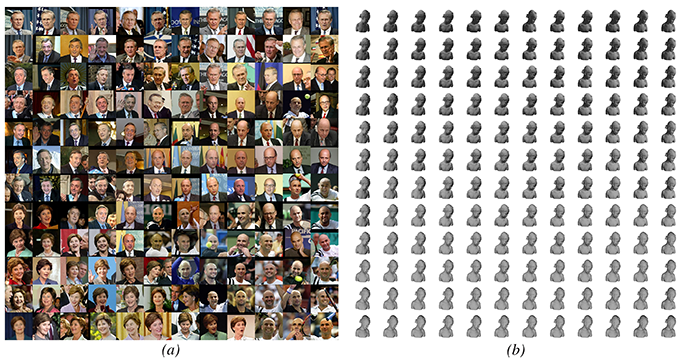}

	\caption{Generating automatic image arrangements with the \SDS~algorithm. (a) Using deep features from a face recognition neural network cluster similar faces together, \eg, bald men (faces are picked at random from the LFW \protect\cite{huang2007labeled} dataset). (b) Automatic image arrangement of images of a 3D model with different lighting. (Images were randomly picked from a 30x30 noisy grid of illumination directions.) Note how the two dimensional lighting direction field is recovered by the \SDS~algorithm: upper-right illuminated model image landed top-right in the grid, and similarly the other corners; images that are placed lower in the layout present more frontal illumination. }
	\label{fig:deepOrdering}
	\vspace{-0.5cm}
\end{figure*}

Our algorithm outperformed \cite{fried2015isomatch} in all experiments (in some cases our algorithm achieved an improvement of more than 50\%). Figures \ref{fig:teaser}, \ref{fig:colors} and \ref{fig:deepOrdering} show some image arrangements from these experiments. Note, for example, how similar faces are clustered together in Figure \ref{fig:deepOrdering} (a), and similar objects are clustered in Figure \ref{fig:teaser} (right).
Further note how the image arrangement in Figure \ref{fig:deepOrdering} (b) nicely recovered the two dimensional parameter of the lighting direction, where Figure~\ref{fig:randomLighting} shows the input random lighting directions renderings of a 3D models.

\paragraph{Coarse-to-fine matching}

\begin{wraptable}[8]{r}{0.4\columnwidth}
        \vspace{-15 pt}\hspace{-20 pt}
        \includegraphics[width=0.43\columnwidth]{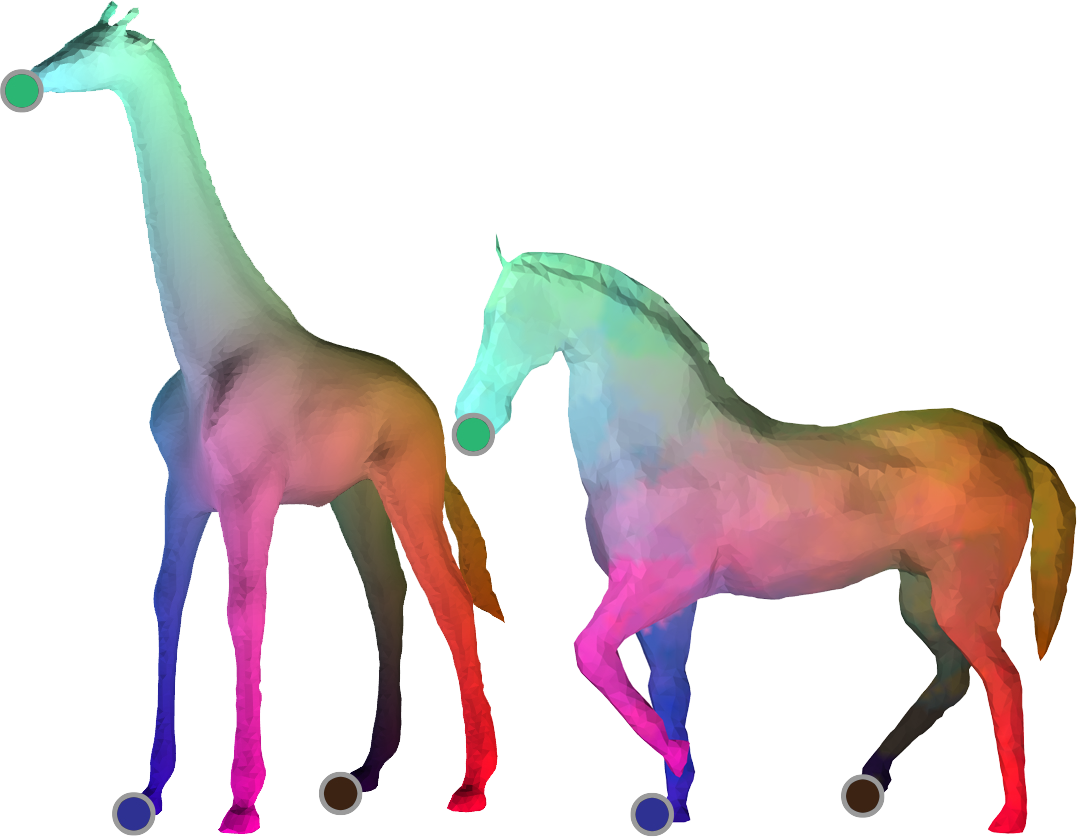}
        \vspace{+5cm}
\end{wraptable}
We consider the problem of matching two shapes $\S$ and $\T$ using sparse correspondences  specified by the user. User input can be especially helpful for highly non-isometric matching problems where semantic knowledge is often necessary for achieving high quality correspondences. The inset shows such example where three points (indicated by colored circles) are used to infer correspondences between a horse and a giraffe.

We assume the user supplied a sparse set of point correspondences,  $\vec{s}_i \to \vec{t}_{i}$, $i=1,\ldots,d $, and the goal is to complete this set to a full correspondence set between the shapes $\S=\set{\vec{s}_1,\ldots,\vec{s}_n} $ and $\T=\set{\vec{t}_1,\ldots,\vec{t}_k}$. Our general strategy is to use a linear term to enforce the user supplied constraints, and a quadratic term to encourage  maps with low distortion.

For a quadratic term we propose a "log-GW" functional. This functional amounts to choosing $p(u,v)=d_L(u,v)^2 $ for the definition of $W$ in \eqref{e:W}, where $d_L $ is a metric on $\RR_+$ defined by
$$d_L(u,v)=\left| \log\frac{u}{v} \right| $$
This metric punishes for high relative distortion between $u,v$, and thus is more suitable for our cause than the standard Euclidean metric used for the GW functional.

As a linear term we propose
$$L(X)=w \brac{ \sum_{i=1}^d (-X_{ii})+  \sum_{i=1}^d \sum_{k,\ell} p \left( d(\vec{s}_i,\vec{s}_q),d(\vec{t}_i,\vec{t}_r) \right) X_{k\ell} }$$
The first summand from the left  penalizes matchings which violate the known correspondences, while the second summand penalizes matchings which cause high distortion of distances to the user supplied points. The parameter $w$ controls the strength of the linear term. In our experiments we chose $w=0.01 \norm{F^TWF} $, where $\norm{F^TWF}=\max\{ |\lambdaMin|,|\lambdaMax| \}$ is the spectral norm of the quadratic form..

\begin{figure}[H]
        \centering
        \includegraphics[width=\columnwidth]{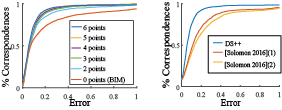}
        \caption{Matching aided by sparse user correspondence. The left graphs shows that our algorithm can exploit user supplied information to outperform state of the art unsupervised methods such as BIM. The right graph shows DS++ outperforms the algorithm of \protect \cite{Justin}  for user aided matching.}
        \label{fig:coarse2fineGraphs}
   
\end{figure}

We applied the algorithm for coarse-to-fine matching on the SHREC dataset  \cite{Shrec},
using $d=3,4,5,6$ of the labeled ground truth points and evaluating the error of the obtained correspondence on the remaining $\ell-d$  points. The number of labeled points $\ell$ is class dependent and varies between $36$ and $7$.

\begin{figure}
        \centering
        \includegraphics[width=\columnwidth]{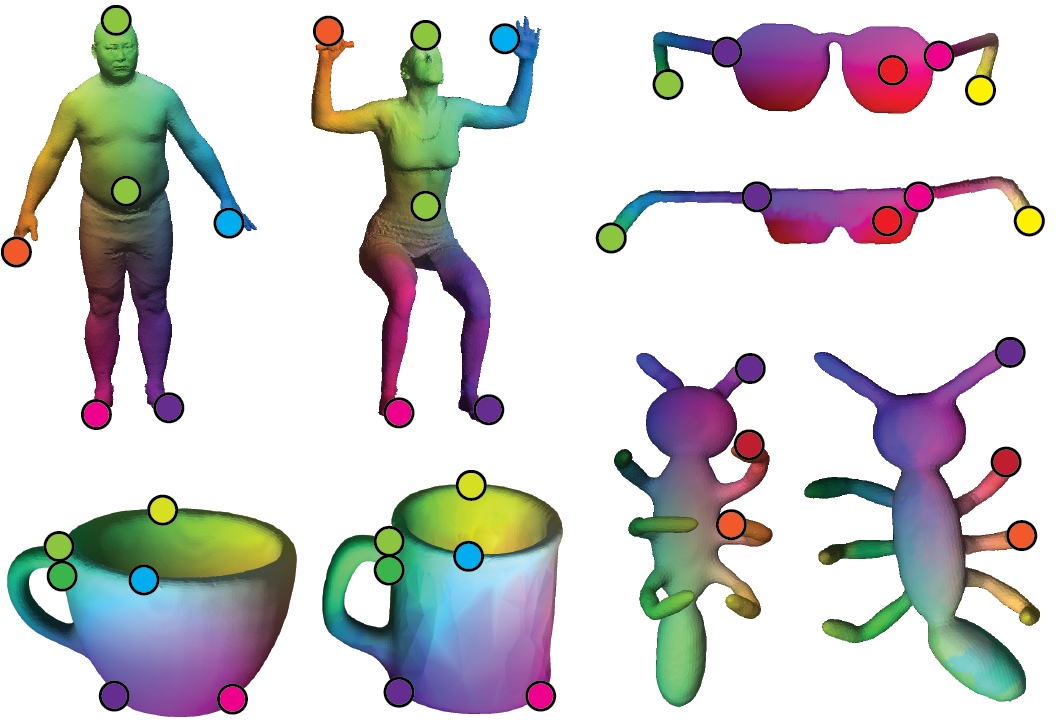}
        \caption{Correspondences obtained using user input. Correspondences were obtained using 6 user input points, with the exception of the correspondence between ants found using only 3 points. Note our method is applicable to surfaces of arbitrary genus such as the genus 1 mugs. \vspace{-0pt} }
        \label{fig:coarse2fine2by2}
\end{figure}

Representative results are shown in Figure~\ref{fig:coarse2fine2by2}.  The graph on the left hand side of Figure~\ref{fig:coarse2fineGraphs} shows that our algorithm is able to use this minimal user supplied information to obtain significantly better results than those obtained by the unaided BIM algorithm \cite{BIM}. The graph compares the algorithms in terms of cumulative error distribution over all 218 SHREC pairs for which BIM results are available.

 The graph on the right hand side of Figure~\ref{fig:coarse2fineGraphs} shows our results outperform  the algorithm presented in \cite{Justin} for matching aided by user supplied correspondences. Both algorithms were supplied with 6 ground truth points.  We ran the algorithm of \cite{Justin}  matching $n=250,k=250$ points (we take $n=k$ since \cite{Justin} does not support injective matching) and using the maximal-coordinate projection they chose to achieve a permutation solution. These results are denoted by (2) in the graph. However we find that better results are achieved when matching only $100$ points, and when using the $L_2$ projection. These results are denoted by (1) in the graph.

\paragraph*{Timing} Typical running times of our optimization algorithm for the energy of \cite{chen2015robust}  matching $n=k=50$ points takes 6 seconds; $n=k=100$  takes 26 seconds; and $n=160$, $k=150$ points takes around 2 minutes (130 seconds). The precomputation of $\lambdaMin$ and $\lambdaMax$ with these parameters requires around 15 seconds, the $L_2$ projection requires 5 seconds, and the remaining time is required for our optimization algorithm.

Parameter values of $n=k=100$ (as well as $n=k=12^2,14^2 $) were used in the image arrangement task from Section~\ref{sec:apps}, and parameter values $n=160,k=150$ were used for our results on the FAUST dataset. For the latter application we also upsampled to $n=450, k=420$ using limited support interpolation and then upsampled to $k=1000, n=5000 $ using greedy interpolation as described in Section~\ref{sec:generalizations}.Limited support interpolation required $117$ seconds and greedy interpolation required $15$ seconds. The total time for this application is around $4.5$ minutes.

The efficiency of our algorithm significantly improves if the product of the quadratic form's matrix $W$ with a vector $x \in \RR^{n^2}$ can be computed efficiently. This is illustrated by the fact that optimization of the sparse functional we construct for the task of resolution improvement with $n=450$ takes similar time as optimization of the non-sparse functional of \cite{chen2015robust} with $n=160$.

 Another case where the product $Wx $ can be computed efficiently is the GW or log-GW energy.  In both cases the product can be computed by multiplication of matrices of size $n\times n$ (see \cite{Justin} for the derivation), thus using $O(n^3) $ operations instead of the $O(n^4) $ operations necessary for general $W$ . Using this energy, matching $n=160 $ points to $k=150$ points takes only 12 seconds, matching $n=270$ points to $k=250 $ points takes 22 seconds, matching $n=500$ to $k=500$ takes 82 seconds, and for $n=k=1,000 $ we require around six minutes (368 seconds).

The efficiency of our algorithm depends linearly on $N$. Minimizing the GW energy with $n=270,k=250 $ using $N+1=5 $ sample takes 12 seconds, approximately half of the time needed when using $N+1=10$ samples. These parameters were used for our results on matching with user input described in Section~\ref{sec:apps}. For this task we also used greedy interpolation to obtain full maps between the shapes, which required an additional 18 seconds. Overall this application required around half a minute.

Our algorithm was implemented on Matlab. All running times were computed using an Intel i7-3970X CPU 3.50 GHz.

\section{Conclusions}
We have introduced the \SDS~algorithm for approximating the minimizer of a general quadratic energy over the set of permutations. Our algorithms contains two components: (i) A quadratic program convex relaxation that is guaranteed to be better than the prevalent doubly stochastic and spectral relaxations; and (ii) A projection procedure that continuously changes the energy to recover a locally optimal permutation, using the convex relaxation as an initialization. We have used recent progress in optimal transport to build an efficient implementation to the algorithm.

The main limitation of our algorithm is that it does not achieve the global minima of the energies optimized. Partially this is unavoidable and due to the computational hardness of our problem. However the experimental results in Figure~\ref{fig:evaluation} show that accuracy can be improved by the SDP method of \cite{Itay} which while computationally demanding, can still be solved in polynomial time. Our future goal is to search for relaxations whose accuracy is close to those of \cite{Itay} but which are also fairly scalable. One concrete direction of research could be finding the `best' quadratic programming relaxation in $O(n^2) $ variables.

\bibliographystyle{acmsiggraph}\vspace{-0pt}
\bibliography{bibdspp}

\appendix
\vspace{-5pt}
\section{Proofs}\label{app:proof}\vspace{-5pt}

\paragraph{Proof of Lemma~\ref{lem:lower_bound}}
The function $f(X)=\fnorm{X}^2 $ is strictly convex and satisfies $f(X)=n $ for all extreme points of $DS$. Therefore
$$E(X,b)-E(X,a) = (a-b)\norm{X}_F^2+(b-a)n \geq 0$$
with equality iff $X$ is a permutation.

\paragraph{Proof of Lemma~\ref{lem:for_thm}}
We omit the proof of the first part of the lemma since it is similar to, and somewhat easier than, the proof of the second part.

To show equivalence of \SDS~with \eqref{e:ourSDP} we show that every minimizer of \SDS~defines a feasible point for \eqref{e:ourSDP} with equal energy and vice versa.

Let $x$ be the minimizer of $E(X,\lambdaMin) $ over the doubly stochastic matrices and let $v$ be the eigenvector of $F^TWF$ of unit Euclidean  norm corresponding to its minimal eigenvalue $\lambdaMin$. Denote $u=Fv$. We define $Y=xx^T+\alpha uu^T$, where we choose $\alpha\geq 0$ so that \eqref{e:SDPtrace} holds. This is possible since $\tr(xx^T) = \norm{X}_F^2\leq n$. Further note that $Y$ also satisfies \eqref{e:SDPSDP} since $\alpha\geq 0$, and \eqref{e:SDPlinearLifted} since
$$AY = Axx^T+\alpha Auu^T = bx^T+\alpha A (Fv)(Fv)^T = bx^T$$
where we used the fact that $Fv$ is a solution to the homogeneous linear equation $Ax=0$. Finally the energy satisfies (ignoring the constant $d$)
\begin{align*}
E(X,Y) &= \tr WY + c^Tx\\ & = x^T W x + \alpha v^T F^TWFv + c^Tx \\
& = x^T W x + \alpha \lambdaMin +c^Tx  \\
&= x^TW x + (n-\fnorm{X}^2)\lambdaMin + c^Tx \\
& = E(X,\lambdaMin)
\end{align*}
Now let $(X,Y) $ be a minimizer of \eqref{e:ourSDP}, we show that $x$ is a feasible solution of our relaxation with the same energy. In fact  due to the previous claim it is sufficient to show that $E(X,\lambdaMin) \leq E(X,Y) $.  The feasibility of $X$ is clear since it is already DS. Next, denote 
$$W_\lambda=W-\lambdaMin I $$
then (ignoring the constant $d$) \begin{align*}
E(X,\lambdaMin) &= \tr W_\lambda xx^T + c^Tx+\lambdaMin n \\ & \overset{(*)}{\leq} \tr W_\lambda Y + c^Tx +\lambdaMin n
\\ & \overset{\eqref{e:SDPtrace}}{=} \tr \ WY + c^T x = E(X,Y)
\end{align*}
The inequality $(*)$ follows from the fact that $A(Y-xx^T)=0 $ due to \eqref{e:SDPlinearEq},\eqref{e:SDPlinearLifted} and therefore since $FF^T $ is the projection onto the kernel of $A$:
$$FF^T(Y-xx^T)=Y-xx^T=(Y-xx^T)FF^T $$
and so $(*)$ follows from
\begin{align*}
\tr W_\lambda(Y-xx^T)&=\tr W_\lambda FF^T(Y-xx^T)FF^T\\
&\tr [F^TW_\lambda F][ F^T(Y-xx^T)F] \geq 0
 \end{align*}
where the last inequality follows from the fact that the two matrices in square brackets are positive semi-definite due to the definition of $\lambdaMin$ and \eqref{e:SDPSDP}.

We now prove the third part of the lemma:
\paragraph*{Comparison with SDP relaxation}
The SDP relaxation of \cite{Itay} is
\begin{subequations} \label{eqn:qam_sdp}
        \begin{align}
        \max _{Y} & \quad \tr{WY}+c^Tx+d \label{eqn:qam_sdp_a} \\
        \mathrm{s.t.} & \quad \tr Y=n \label{eqn:qam_sdp_d} \\
        & \quad x \geq 0 \label{eqn:qam_sdp_ineq}\\
        & \quad Ax=b  \label{eqn:qam_sdp_c} \\
        & \quad Y \succeq xx^T \label{eqn:qam_sdp_b} \\
        & \quad Y \geq0 \label{eqn:qam_sdp_e} \\
        & \quad \sum_{\ik\il\ir\is} Y_{\ik\il\ir\is}  = n^2 \label{eqn:qam_sdp_f} \\
        & \quad Y_{\ik\il\ir\is} \leq \begin{cases}
        0, & \mathrm{if} \ \ \ik=\ir,\  \il\neq\is \\
        0, & \mathrm{if} \ \ \il=\is,\  \ik\neq\ir \\
        \min\set{X_{\ik\il}, X_{\ir\is}}, & \mathrm{otherwise} \end{cases}  \label{eqn:qam_sdp_g}
        \end{align}
\end{subequations}
where $Y_{qrst} $ is the entry replacing the quadratic  monomial $X_{qr}X_{st} $.
We note this relaxation contains all constraints from the SDP relaxation \eqref{e:ourSDP} with the exception of \eqref{e:SDPlinearLifted}. It also contains the  additional constraints \eqref{eqn:qam_sdp_e}-\eqref{eqn:qam_sdp_g}  which do not appear in  \eqref{e:ourSDP}. Thus to show that \cite{Itay} is tighter than our relaxation it is sufficient to show that \eqref{e:SDPlinearLifted} is implied by the other constraints of \cite{Itay}. We recall that \eqref{e:SDPlinearLifted} represent all constraints obtained by multiplying linear equality constraints by a linear monomial.

For a quadratic polynomial
$$g(x)=x^TWx+c^Tx+e $$
let us denote by $\bar g(x,Y)$ the linearized polynomial
$$\bar g(x,Y)=\tr{WY}+c^Tx+e $$
We will use the following property of SDP relaxations (see \cite{PMexact}):
If a quadratic polynomial $g $ is of the form $g=p^2 $ then
\begin{enumerate}
        \item For any feasible $x,Y $ we have $\bar g(x,Y) \geq 0 $.
        \item If $\bar g(x,Y)=0 $ is satisfied for all feasible $x,Y $, then for any quadratic $f$ of the form $f=pq$ we have
        $\bar f(x,Y)=0$.
\end{enumerate}

Accordingly, it is sufficient to show that the squares $g_q=p_q^2, h_r=m_r^2 $ of all the linear equality polynomials
$$p_q(X)=\sum_r X_{qr}-1 \quad , \quad m_r(X)= \sum_r X_{qr}-1  $$
satisfy $\bar g_q=0, \bar h_r=0 $. We obtain $\bar g_q=0 $ from
\begin{align*}
0 &\leq \bar g_q(x,Y)=\sum_{r,t} Y_{qrqt}-2\sum_r X_{qr}+1 \\
  &\leq^{\eqref{eqn:qam_sdp_g}} \sum_r X_{qr}-2\sum_r X_{qr}+1=0
\end{align*}
the proof that $\bar h_r=0$ is identical.

\section{Sparsity pattern for improving matching resolution} We construct a sparsity pattern for the task of matching $\s_1,\ldots,\s_k $ to $\t_1,\ldots,\t_n $ using known correspondences $\hat{\s}_\ell \mapsto \hat{\t}_\ell, \ell=1,\ldots,r  $.

For each $\s_i$ we use the following procedure to determine which correspondence will be forbidden: We find the five  matched points $\hat{\s}_{\ell_1},\ldots, \hat{\s}_{\ell_5} $ which are closest to $\s_i$ and compute the geodesic distance of these points from $\s_i$. This gives us a feature vector $v \in \RR^5 $. We then compute the geodesic distances of each of the points $\t_j, j=1,\ldots,n $ from the matched points $\hat {\t}_{\ell_1},\ldots \hat{\t}_{\ell_5} $ corresponding to the five closets points to $\s_i$. This gives us $n$ feature vectors $v_j \in \RR^5 $. For $\t_j$ to be a viable match we require that $\norm{v_j-v}_2 $ be small. We therefore allow the top $20\%$ of the correspondences according to this criteria.

To symmetrize this process, we use the same procedure to find permissible matches for each $\t_j$, and then select as permissible all matches $\s_i \mapsto \t_j$ which were found permissible either when starting from $\s_i$ or when starting from $\t_j$.

\end{document}